\documentstyle[aps,epsfig,prl,multicol]{revtex}
\begin{document}
\draft

\title{Amplitude and Gradient Scattering in Waveguides with Corrugated Surfaces}

\author {F. M. Izrailev$^1$, G. A. Luna-Acosta$^1$, J. A. M\'endez-Berm\'udez$^1$, and M. Rend\'on$^2$}
\address{$^1$Instituto de F\'{\i}sica, Universidad Aut\'onoma de Puebla,
Apdo. Postal J - 48, Puebla 72570, Mexico. \\
$^2$ Facultad de Ciencias de la Electr\'onica,
Universidad Aut\'onoma de Puebla, Puebla, Pue., 72570, Mexico.}

\date{\today}
\maketitle
\begin{abstract}

\end{abstract}

\begin{abstract}
We study chaotic properties of eigenstates for periodic quasi-1D waveguides 
with "regular" and "random" surfaces. Main attention is paid to the role of 
the so-called "gradient scattering" which is due to large gradients in 
the scattering walls. We demonstrate numerically and explain theoretically
that the gradient scattering can be quite strong even if the amplitude of
scattering profiles is very small in comparison with the width of waveguides.
\end{abstract}


During last decade much attention has been paid to the theory of
quasi-1D disordered solids with the so-called {\it bulk scattering}.
By this term one describes the situation where the whole
volume of a scattering region contains scatters whose density determines the
mean free path $\lambda$ for a propagation of electrons. According to the theory,
apart from $\lambda$, transport properties of finite samples are described by 
two other characteristic lengths: size $L$ of a sample and {\it localization
length}\, $l_{\infty}$. The latter is deterimied by the degree of the decrease of the amplitude
of eigenstates along infinite samples with the same scattering characteristics.
The core of the modern theory of the transport for such quasi-1D systems
is the so-called {\it single-parameter} scaling. It was shown that when
the mean free path is much less than both $L$ and $l_{\infty}$, {\it all} statistical
characteristics of the transport are fully described by the only scaling parameter which is
the ratio of the localization length to the size of a sample 
(see, e.g. \cite{sigma} and references therein). 

Another kind of quasi-1D systems that has attracted much attention in past few years,
is the many-mode waveguide with rough surfaces. In this case the scattering is
entirely related to statistical characteristics of scattering walls, therefore,
one can speak about {\it surface scattering}. For some time it was believed 
that the surface scattering can be analytically described by modified methods 
thoroughly developed for bulk scattering. However, recent numerical studies of
such systems \cite{recent,SFYM99} have revealed a principal difference between 
surface and bulk scattering (see discussion 
and references in \cite{LFL98}). Specifically, it was found that the transport 
through quasi-1D waveguides 
with rough surfaces essentially depends on many characteristic lengths, not on one length
as in the case of the bulk scattering. This fact is due to a non-isotropic 
character of scattering in the channel space. In particular, the transmission 
coefficient smoothly decreases with an increase of the angle of incoming waves, since
characterictic lengths for backscattering are different for different
channels \cite{SFYM99,IM03}.

The latter subject of the surface scattering has a direct link to the 
problem of {\it quantum chaos}. The point is that the waveguides with rough
walls can be treated from the viewpoint of classical and quantum mechanics
that describe a particle moving inside billiards and having multiple reflections
from the walls. One of the problems of quantum chaos is the quantum-
classical correspondence for the situation when, in the classical limit,
global properties of the motion of a particle are strongly chaotic. More
specifically, it is of great interest to find what is the fingerprint of
classical chaos in quantum eigenstates of closed/periodic billiards, 
as well as the relation of statistical properties of the transport through
open billiards to the underlying classical chaos.

In this paper we investigate the properties of quantum eigenstates
of billiards with regular and rough walls, with the
application to the wave scattering through quasi-1D waveguides
with surface scattering. To be specific,
we consider quasi-1D waveguides that are periodic in the  
$x-$direction with period $2\pi$. The upper and low walls 
are given by the functions, $f_1(x)=f_1(x+2\pi) 
= d + a_1 \xi_1 (x)$
and $f_2(x)=f_2(x+2\pi) = a_2 \xi_2 (x)$ where 
$d$ is the average width of the waveguides and
$a_{1,2} \ll d$ stands for the amplitude of the scattering walls.
Our interest is in the structure of eigenstates
of the corresponding Hamiltonian with zero boundary conditions
on the two walls. 

For our purpose it is convenient to pass to the variables
$u = x,\  v = \frac{f_2(x)-y}{f_2(x)-f_1(x)}$
in which the new Hamiltonian
$\hat H = \hat H^0 + \hat V$ describes a particle moving inside 
a waveguide with flat boundaries in new $u, v$ coordinates 
(see details in \cite{qrch}). 
Here $\hat H^0=\frac{1}{2m_e}(\hat P_u^2 + \hat P_v^2)$
and the effective potential $\hat V(u,v, \hat P_u, \hat P_v)$
depends on the functions $f_1$ and $f_2$, with $\hat P_u, \hat P_v$
as the canonical momentums. 
The solution of the Schr\"odinger
equation for $\hat H$ can be written in the form $\psi_E(u,v)=\exp
(i\chi)\,\psi_\chi(u,v)$. Since statistical properties of eigenstates
of $\hat H$ do not depend on specific value of the Bloch index $\chi$
inside the first Brillouin band, all numerical data were obtained for a
specific value of $\chi$. By
expanding $\psi_\chi(u,v)$ in the basis of $\hat H_0$, 
one can find the matrix representaion of $\hat H$ in the unperturbed basis
specified by the two indexes $n$ and $m$ \cite{one}. 

The Hamiltonian matrix elements are given by
\begin{eqnarray}
H^k_{mnm'n'} & = & \frac{\hbar^2}{2\pi} \left\{ \pi (n+k)^2 \delta_{nn'} \delta_{mm'} + 
\left[ \frac{m^2\pi^2}{2} \left( J^2_{n'n} + J^3_{n'n} + J^6_{n'n} \right) + \left( \frac{1}{8} + 
\frac{m^2\pi^2}{6} \right) J^4_{n'n} \right] \delta_{mm'} \right. \nonumber \\
& + & i (n+n'+2k) \frac{mm'}{m^2-m'^2} \left[ J^1_{n'n} - (-1)^{m+m'} 
\left( J^1_{n'n} + J^5_{n'n} \right) \right] \\ 
& + & \left. \frac{2mm'(m^2+m'^2)}{(m^2-m'^2)^2} \left[ -J^6_{n'n} + (-1)^{m+m'} 
\left( J^6_{n'n} + J^4_{n'n} \right) \right] \right\} \nonumber 
\end{eqnarray}
where

\begin{equation}
J^1_{n'n} = \int^{2\pi}_0 \frac{f_2'}{f_1-f_2} \,\, e^{i(n'-n)u} du \,;
\,\,\,\,\, 
J^2_{n'n} = \int^{2\pi}_0 \frac{(f_2')^2}{(f_1-f_2)^2} \,\, e^{i(n'-n)u} du \,,
\end{equation}

\begin{equation}
J^3_{n'n} = \int^{2\pi}_0 \frac{1}{(f_1-f_2)^2} e^{i(n'-n)u} du  \,; 
\,\,\,\,\,
J^4_{n'n} = \int^{2\pi}_0 \frac{(f_1'-f_2')^2}{(f_1-f_2)^2} e^{i(n'-n)u} du \,,
\end{equation}

\begin{equation}
J^5_{n'n} = \int^{2\pi}_0 \frac{f_1'-f_2'}{f_1-f_2} e^{i(n'-n)u} du \,;
\,\,\,\,\,
J^6_{n'n} = \int^{2\pi}_0 \frac{f_2'(f_1'-f_2')}{(f_1-f_2)^2} e^{i(n'-n)u} du .
\end{equation}
with $f_{1,2}' \equiv \partial f_{1,2}(u) /\partial u$.\\

One should note that the matrix elements in the new variables $u, v$ depends both on
$f_{1,2}$ and on their derivatives $f_{1,2}'$. This very
fact demonstrates a highly non-trivial role of scattering walls since it is
a problem to separate the influence of the {\it amplitude scattering} from the
{\it gradient scattering}. As was shown in Ref.\cite{mak1}, the perturbation
approach is quite complicated and should be performed in 
a special way.

Our goal is two-fold. First, we would like to compare the case of a
"regular" upper profile $\xi_1(u) = \cos(u)$ that has been studied in
detail in \cite{qrch,one} and \cite{others}, 
with the "random" one $\xi_1(u) = \sum_N^{N_T} A_N \cos(Nu)$ where the 
amplitudes $A_N$ are chosen at random and $N_T=100$.
 In both these cases we assume $\xi_2(u)=0$ and we refer
to this case as to the "symmetric case". In this way we explore the
role of roughness of the surface scattering.

Second, we wish to analyse the role of the gradient scattering that is due to
the derivatives $f_{1,2}'$. For this we consider the "asymmetric case"
with $f_1 = d + a \xi (u)$ and $f_2 = a \xi (u)$ with $a=a_1=a_2,\, \xi (u)=\xi_1(u)$. 
As one can see from the
expressions for the matrix elements, in this case the scattering is only due
to the "gradient terms" since $f_1-f_2=d=const$. Specifically, we have 
$J^4_{n'n}=J^5_{n'n}=J^6_{n'n}=0$, and the rest of the 
integrals are reduced to 

\begin{equation}
J^1_{n'n} = \epsilon \int^{2\pi}_0 \xi_u e^{i(n'-n)u} du \,;\,\,
J^2_{n'n} = \epsilon^2 \int^{2\pi}_0  \xi_u^2 e^{i(n'-n)u} du \,;\,\,
J^3_{n'n} = \frac{1}{d^2} \int^{2\pi}_0  e^{i(n'-n)u} du \,\,.
\end{equation}
where $\xi_u \equiv \partial \xi(x) /\partial u$, and $\epsilon \equiv a/d$.\\

One natural representation of the Hamiltonian matrix
$H_{l,l'}(\chi)= <l\mid\hat H\mid l'>_\chi$ is the
"channel representation" for which one fixes the values of $n$
starting from the lowest one, $n=-N_{max}$, and running over all values of $m$. 
In numerical simulations we have to make a cutoff for the values
of $m$ and $n$ in the Hamiltonian matrix. 
Our data refer to the ranges, $1 \leq m
\leq M_{max}$ and $|n| \leq N_{max}$ with $N_{max}=32
\,,M_{max}=62$, for which the total size of the Hamiltonian matrix is
$L=(2N_{max}+1)M_{max}=4030$.  

>From the structure of the Hamiltonian matrices shown in Fig.1 one can make
some important conclusions. First, by passing from "regular" wavegudes with $N_T=1$
to "random" ones with $N_T=100$ in both symmetric and asymmetric cases, the Hamiltonian
matrices tend to be fully filled by off-diagonal elements. However, 
the matrices correponding to the random
waveguides keep the block structure, thus indicating some regularity in spite of
a completely random character of the surface scattering. This fact was shown to result in
a kind of "non-ergodicity" for the eigenstates of the total Hamiltonian $\hat H$ for any
high energy, see details in Ref. \cite{ournew}. It is clear that standard theoretical
approaches based on completely random matrices can not adequately discribe the
structure of eigenstates. As was shown in Ref.\cite{IM03}, for open waveguides with
rough surfaces and large number of channels there are many characteristic lengths, in
contrast to the bulk scattering where the so-called single-parameter scaling holds.

First, Figure 1 clearly demonstrates that the eigenstates of $\hat H$
are expected to be much more extended in the unperturbed basis for rough profiles, 
than for one-cosine profiles. Second, it is instructive to make a comparison between the
symmetric and asymmetric cases (compare (a) with (b) and (c) with (d) in Fig.1). The data show
that the matrices for the assymetric cases have smaller elements than those for the symmetric ones.
This is in correspondence with the fact that large part of terms in the expressions for 
the matrix elements of $\hat H$ vanishes due to the absence of the amplitude scattering.
Therefore, one can expect that the eigenstates are less random for the asymmetric case.

\begin{figure}[htb]
\begin{center}
\epsfig{file=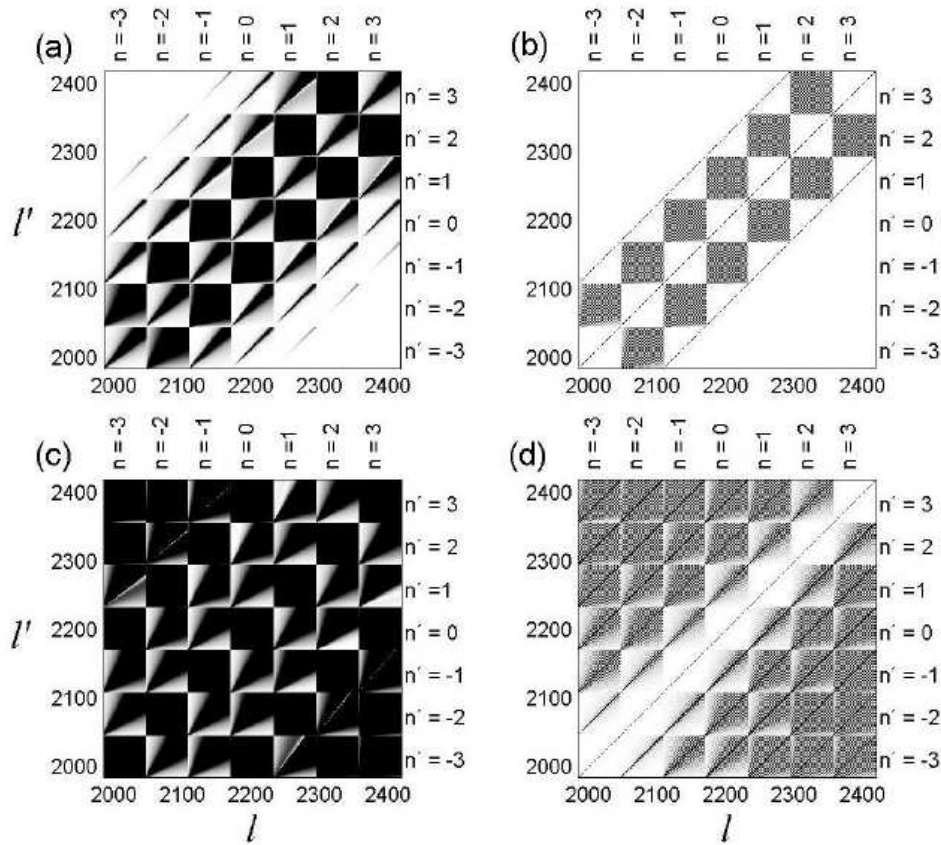,width=5in}
\vspace{.5cm}
\caption{Central part of the Hamiltonian matrix. The $62 \times 62$
blocks corresponding to $n,n'=[-3,3]$ are shown. The larger the amplitudes of the matrix elements the more black the corresponding regions are. Four cases
refer to: (a) one-cosine symmetric waveguide, (b) one-cosine asymmetric waveguide, (c) symmetric waveguide with a rough wall, $N_T=100$, (d) asymmetric
waveguide with rough walls, $N_T=100$.}
\end{center}
\end{figure}

In order to analyze the structure of eigenstates of $\hat H$ in detail, we have
diagonalized the Hamiltonian matrices shown in Fig.1 and constructed the
"state matrices" $|C^\alpha_l|^2$. Here $C^\alpha_l$ are the amplitudes of the
eigenstates in the basis representaion given by the index $l$. Namely, the index $l$
refers to unperturbed basis states that correspond to the unperturbed Hamiltonian
$\hat H^0$. The index $\alpha$ refers to a specific exact eigenstate. 
All eigenstes are reordered in increasing energy, 
with $\alpha=1$ the ground state. Therefore, to understand how
strongly localized/extended are the exact eigenstates in the unperturbed basis, one should
fix the value of $\alpha$ and explore the dependence of $|C^\alpha_l|^2$ on $l$. 

Let us now compare the symmetric cases with the asymmetric ones, see Fig.2. The most
important conclusion that can be deduced from the data is that the eigenstates are,
in general, more extended in the asymmetric cases. Specifically, for the same small values of
$\alpha$ there are more components $C^\alpha_l$ 
with large values of $l$, in comparison with
the symmetric case, compare (c) with (d). 
At a first glance, this looks strange since for asymmetric
profiles the Hamiltonian matrices are obviously less "random" than for the
symmetric ones, as is mentioned above. 
Close inspection of the data in Fig.2 shows that there is an additional
effect that is also important in connection with the structure of eigenstates. Namely,
the eigenstates for asymmetrical cases (b) and (d) turn out to be more {\it sparse} in
comparison with the cases (a) and (c). This is manifested by a large number of "holes" 
along each line in (b) and (d) for fixed values of $\alpha$, in the comparison with the
cases (a) and (c). 
Thus, the eigenstates for waveguides with only gradient scattering (asymmetric case)
are more extended in the unperturbed basis, and, at the same time, more sparse than
for the wavegudies with both gradient and amplitude scattering (symmetric case). 

This phenomenon is important in view of the scattering properties through
waveguides of {\it finite} size with the profiles such as considered here. 
As is known, chaotic structure of
eigenstates of closed (or periodic) waveguides/billiards is directly related 
to the scattering
properties of open systems with the same profiles. For example, the degree of localization
of eigenstates of closed systems determines the degree of localization of scattering states,
and correspondingly, the value of transmission through open systems. 

\begin{figure}[htb]
\begin{center}
\epsfig{file=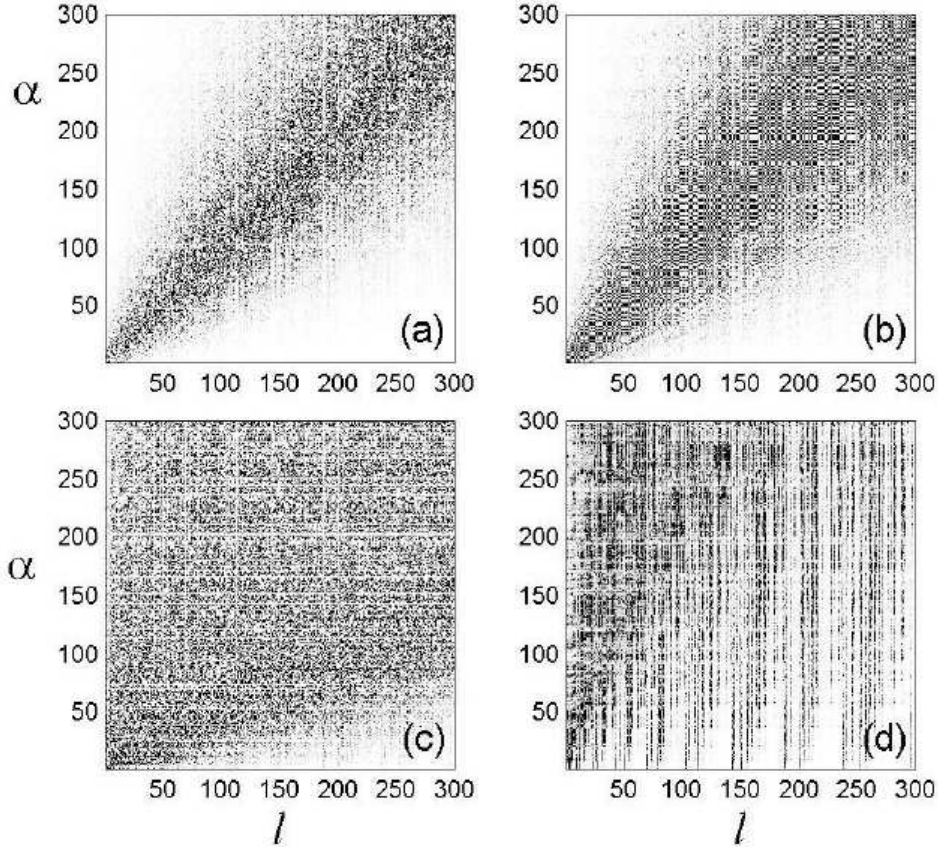,width=5in}
\vspace{.5cm}
\caption{Lower part of the state matrix $|C^\alpha_l|^2$. The data for all cases (a-d) refer to the same profiles for which the structure of Hamiltonian matrices is shown in Fig.1. Here darker regions correspond to larger
values of $|C^\alpha_l|^2$.}
\end{center}
\end{figure}

In conclusion, we have studied the structure of eigenstates of quasi-1D periodic waveguides
with regular and random walls. Main attention was paid to the role of gradient
scattering in comparison with the amplitude scattering. It was shown that for the case when
the scattering walls have large number of harmonics, the gradient scattering itself is
very strong. This is demonstrated by the data obtained for the waveguides with asymmetric
walls, for which the amplitude scattering is absent. It was also
revealed that the role of the gradient scattering is highly non-trivial. Specifically,
the gradient scattering turns out to be relatively strong in the absence of the amplitude
scattering. \\

We are grateful to N. Makarov for fruitful discussions. This work was 
supported by the CONACyT (Mexico) Grant No. 34668-E, and IIG3G02, VIEP, BUAP.

\end{document}